 \documentclass[final,3p,times,twocolumn]{elsarticle}

\usepackage{amssymb}

\biboptions{sort&compress}

\journal{NIM B}

\begin{document}

\begin{frontmatter}



\title{Activation cross-sections of longer lived products of proton induced nuclear reactions on manganese up to 70 MeV}


\author[1]{F. Ditr\'oi\corref{*}}
\author[1]{F. T\'ark\'anyi}
\author[1]{S. Tak\'acs}
\author[2]{A. Hermanne}
\author[3]{H. Yamazaki} 
\author[3]{M. Baba}
\author[3]{A. Mohammadi} 
\cortext[*]{Corresponding author: ditroi@atomki.hu}

\address[1]{Institute of Nuclear Research of the Hungarian Academy of Sciences (ATOMKI),  Debrecen, Hungary}
\address[2]{Cyclotron Laboratory, Vrije Universiteit Brussel (VUB), Brussels, Belgium}
\address[3]{Cyclotron Radioisotope Center (CYRIC), Tohoku University, Sendai, Japan}

\begin{abstract}
In the frame of a systematic study of the activation cross-sections of the proton induced nuclear reactions, excitation functions of the $^{55}$Mn(p,x)$^{154,152g}$Mn, $^{51}$Cr and $^{48}$V were measured up to 70 MeV. Cross-sections were measured with the activation method using a stacked foil irradiation technique and high resolution  $\gamma$-ray spectrometry. The experimental data are analyzed and compared to the earlier results and to the prediction  of the EMPIRE-3 as well as the TALYS theoretical model code in the TENDL-2012 library. From the measured cross-section data integral production yields were calculated. Practical applications of the cross-sections are discussed. 
\end{abstract}


\end{frontmatter}


\section{Introduction}
\label{1}
Manganese is an important alloying additive for ferrous and non-ferrous alloys imparting many beneficial properties. Structural components of different technological units are made of stainless steel containing around 2\% of manganese. Manganese metal is also a key component of aluminum copper alloys. These alloys are widely used in accelerator technology where the reliable estimation of the activation levels is important. The $^{54}$Mn activation product has proper decay characteristics for application in thin layer activation technology for investigation wear, erosion and corrosion processes for units made from alloys containing manganese. The study has also its importance in the frame of our systematic investigation for improvement of the theoretical codes for proton induced reactions, having presently only moderate performance for estimation of cross-sections for activation products far from the mass of the target nuclei. The element manganese has only one stable isotope $^{55}$Mn, in such a way isotopic cross-sections can be obtained by irradiating manganese with natural isotopic composition.

\section{Experimental}
\label{2}
As a part of a series of systematic studies \cite{1,2,3,4}, excitation functions were measured by activation, using stacked foil irradiation technique and high resolution  $\gamma$-spectrometry to determine the amount of the produced radioisotopes. The experimental circumstances are summarized in the Table 1. The target composition gives also the possibility to determine the elemental cross-section of the $^{62,65}$Zn produced on the Cu content of the target from the same experiment, simultaneously to the cross-section on manganese, which will be published elsewhere, together with the results of other irradiations on copper target.
The decay and spectrometric characteristics of the activation products were taken from the NUDAT2 data base \cite{5} and are summarized in Table 1 \cite{6}, together with the Q-value of the contributing reactions \cite{7}. 
The cross-sections are relative to the recommended cross-sections of the used monitor reactions. The excitation function of the monitor reactions were measured simultaneously in the whole energy range covered. The measured excitation functions of the monitor reactions are shown in Fig. 1 and Fig. 2. The simultaneously measured excitation functions of the monitor reactions allow to control the number of the incident particles and the energy degradation through the stack \cite{8}. The activities of the irradiated targets were measured without chemical separation by high resolution $\gamma$-ray spectrometry. The decay of the investigated isotopes was followed by counting each foils several times (2-3 measurements), starting from a few hours after EOB (end of bombardment) until weeks later. The cross-sections and their uncertainties were determined by using the widely used activation and the decay formulas \cite{9}. The decay data of the investigated isotopes and the Q-values of the contributing reactions were taken from NUDAT \cite{10} and from the Q-value calculator of NNDC \cite{7} (see Table 1). The uncertainty of the measured values was determined in a standard way \cite{11}: the independent relative errors of the linearly contributing processes (number of the bombarding particles (8\%), number of the target nuclei (5\%), decay data (3\%), detector efficiency (5\%) and peak area (1-7\%)) were summed quadratically and the square root of the sum was taken. The uncertainties of the non-linear processes like half-life, irradiation time and measuring time were neglected. The uncertainty of the primary beam energy was estimated to be around $\pm$0.3 MeV. The uncertainty of the beam energy increases along the stack due to cumulative straggling effect and the uncertainty on the determination of the foil thickness in the stack and the estimated inhomogeneity. The nuclear data used for the data evaluation are summarized in Table 2.

\begin{figure}[h]
\includegraphics[scale=0.3]{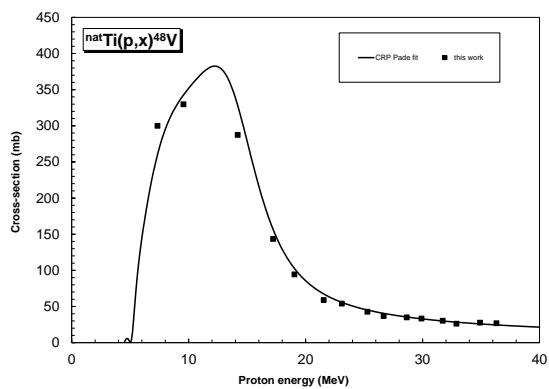}
\caption{Excitation function of the $^{nat}$Ti(p,x)$^{48}$V monitor reaction compared with the recommended values}
\end{figure}

\begin{figure}[h]
\includegraphics[scale=0.3]{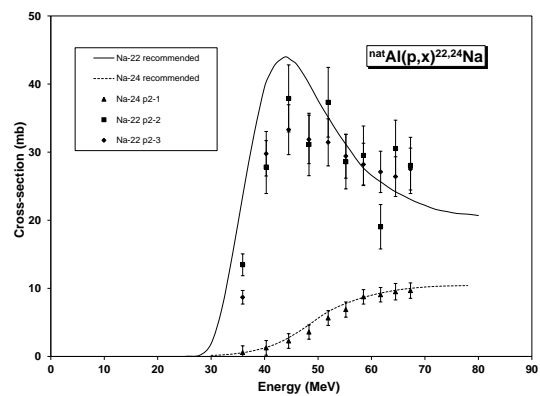}
\caption{Excitation functions of the $^{27}$Al(p,x)$^{22,24}$Na  monitor reactions compared with the recommended values}
\end{figure}

\begin{table*}[t]
\small
\caption{Main experimental parameters}
\centering
\begin{center}
\begin{tabular}{|p{1.3in}|p{1.5in}|p{1.5in}|p{1.5in}|} \hline 
\textbf{Laboratory:} & \textbf{VUB, Brussels} & \textbf{CYRIC, Sendai} & \textbf{CYRIC2, Sendai} \\ \hline 
Incident particle & Proton  & Proton & Proton \\ \hline 
Method  & Stacked foil & Stacked foil & Stacked foil \\ \hline 
Target composition\newline and thickness  & Cu(86)Mn(12)Ni(2) alloy foil,\newline Goodfellow $>$99.98\%,\newline 25 $\mu$m & Cu(86)Mn(12)Ni(2) alloy foil 25 $\mu$m, Goodfellow $>$99.98\%\newline  & Mn foil 10 $\mu$m \newline Goodfellow $>$99.98\%\textbf{} \\ \hline 
Number of target foils & 17 & 12, 150 (2 stacks) &  \\ \hline 
Stack composition & MoRe(50 $\mu$m), \newline Ti(12 $\mu$m), \newline CuMnNi(25 $\mu$m), \newline Al(100,200 $\mu$m),\newline  Mo(12 $\mu$m) & CuMnNi(25 $\mu$m),\newline MoRe(50 $\mu$m),\newline Al(50, 100, 200, 500 $\mu$m),\newline ScO (350 $\mu$m)\textbf{} & Zr(100 $\mu$m)\newline Rh(12 $\mu$m)\newline Mn(10 $\mu$m)\newline Ag(50 $\mu$m)\newline Polyethilene backing\textbf{} \\ \hline 
Accelerator & CGR 560 cyclotron of Vrije Universiteit Brussels\newline (VUB) & AVF-930 cyclotron of the Cycloton Laboratory (CYRIC) of the Tohoku University & AVF-930 cyclotron of the Cycloton Laboratory (CYRIC) of the Tohoku University \\ \hline 
Primary energy & 37 MeV & 70 MeV & 70 MeV \\ \hline 
Covered energy range & 37-0 MeV & 70-35 MeV & 70-23 MeV \\ \hline 
Irradiation time & 71 min & 20 min & 1 h\textbf{} \\ \hline 
Beam current & 61 nA & 26 nA  & 100 nA\textbf{} \\ \hline 
Monitor reaction, \newline [recommended values]  & ${}^{nat}$Ti(p,x)${}^{48}$V reaction [1] & ${}^{27}$Al(p,x)${}^{22,}$${}^{24}$Na  reaction [1] & ${}^{27}$Al(p,x)${}^{22,}$${}^{24}$Na  reaction \\ \hline 
Monitor target and thickness & ${}^{nat}$Ti, 12 $\mu$m & ${}^{nat}$Al, 50, 100, 200 $\mu$m & ${}^{nat}$Al, 50, 100, 500 $\mu$m \\ \hline 
detector & HpGe & HpGe & HpGe \\ \hline 
$\gamma$-spectra measurements & 4 series & 3 series & 2 series\textbf{} \\ \hline 
 Cooling times\newline (Source --detector distances) & 2 h (45 cm), 25 h (20 cm), 90 h (5 cm), 700 h(5 cm) & 45h (5 cm), 170h (5 cm), 1000h (5 cm) & 48-168 h, \\ \hline 
\end{tabular}
\end{center}
\end{table*}

\begin{table*}[t]
\small
\caption{Decay characteristics of the investigated activation products}
\centering
\begin{center}
\begin{tabular}{|p{0.6in}|p{0.7in}|p{0.8in}|p{0.8in}|p{1.4in}|p{0.8in}|} \hline 
\textbf{Nuclide} & \textbf{Half-life} & \textbf{E${}_{\gamma}$(keV)} & \textbf{I${}_{\gamma }$(\%)} & \textbf{Contributing reactions} & \textbf{Q-value (keV)} \\ \hline 
\multicolumn{6}{|c|}{${}^{55}$Mn(p,x)${}^{54,52}$Mn,${}^{51}$Cr,${}^{48}$V} \\ \hline 
${}^{54}$Mn & 312.05 d & 834.848 & 99.976 & ${}^{55}$Mn(p,pn) & ~-10226.53 \\ \hline 
${}^{52g}$Mn & 5.591 d  & 744.233\newline 935.544\newline 1434.092 & 90.0\newline 94.5\newline 100.0 & ${}^{55}$Mn(p,p3n)\newline ${}^{52}$Fe decay & -31219.09\newline -34375.27 \\ \hline 
${}^{51}$Cr & 27.701 d & 320.0824 & 9.910 & ${}^{55}$Mn(p,2p3n) & ~-37764.7 \\ \hline 
${}^{48}$V & 15.9735 d & ~983.525\newline 1312.106 & 99.98\newline ~98.2 & ${}^{55}$Mn(p,3p5n)\newline ${}^{48}$Cr decay & -68169.73\newline -70608.29 \\ \hline 
\multicolumn{6}{|c|}{${}^{27}$Al(p,x)${}^{22,24}$Na, ${}^{nat}$Ti(p,x)${}^{48}$V} \\ \hline 
${}^{22}$Na & 2.6027 y & 1274.537 & 99.941 &  &  \\ \hline 
${}^{24}$Na & 14.997 h & 1368.626 & 99.9936 &  &  \\ \hline 
${}^{48}$V & 15.9735 d & ~983.525\newline 1312.106 & 99.98\newline ~98.2 &  &  \\ \hline 
\end{tabular}

\end{center}
\begin{flushleft}
\footnotesize{When complex particles are emitted instead of individual protons and neutrons the Q-values have to be decreased by the respective binding energies.\\

\noindent Increase Q-values if compound particles are emitted: np-d, +2.2 MeV; 2np-t, +8.48 MeV; n2p-${}^{3}$He, +7.72 MeV; 2n2p-$\alpha$, +28.30 MeV}
\end{flushleft}
\end{table*}

\section{Tehoretical calculations}
\label{3}
In order to estimate the predicted values of the cross-sections both TALYS \cite{12} (by using the TENDL-2012 library \cite{13}) and the latest version of EMPIRE-3 \cite{14,15} theoretical model codes were used. For the EMPIRE calculations, the optical potential parameters were taken from the Recommended Input Parameter Library RIPL-2 \cite{16}. The TENDL library was constructed by using the TALYS nuclear model code and the RIPL-2 [16] parameters were applied for the blind calculations. The quality of the models in the TALYS code was improved starting from phenomenological default parameters for TALYS 1.0 and more microscopic options in TALYS 1.2 version \cite{17}. In TENDL 2012 significant improvement is observed for the predictive power for the activation cross-sections of proton induced reactions. 
The TALYS and EMPIRE-3 codes calculate the population of different low-lying levels and can also estimate isomeric ratios for these levels. For all identified radioisotope the EMPIRE-3.1 (Rivoli) code was used \cite{14,15}. As far as the charged particle induced reaction this version of EMPIRE code contains the following new features: RIPL-3 library of input parameters; new version of Coupled Channel code ECIS-2006; coupled Channel code OPTMAN for soft-rotor calculations; parity dependent level densities; new parameterization of EGSM level densities; three additional ejectiles (d, t, $^3$He). The first run of the EMPIRE-3 code was performed with FITLEV$\>$0 parameter in order to fine tune the fit on the experimental level densities.

\section{Results}
\label{4}
Being the manganese monoisotopic ($^{55}$Mn) all investigated reactions result in isotopic cross-sections. The outgoing particles could be from light (p,n,d) to more complex particles (t,$^3$He,$\alpha$). 

\subsection{The $^{55}$Mn(p,x)$^{54}$Mn reaction}
\label{4.1}
The long lived radioisotope $^{54}$Mn (T$_{1/2}$ = 312.2 d) can exclusively be produced by (p,np) reaction, where a complex particle (d) can also be emitted instead of the np particle group.  The results compared with the TALYS and EMPIRE-3 calculations are presented in Fig. 3. As it is seen from the figure, our new results are in good agreement with the earlier values of Stueck \cite{18}, Michel \cite{19,20}, Levkovskij \cite{21}, Cohen \cite{22} and Gusakow \cite{23} in a large part of the covered energy range. TENDL-2012 and EMPIRE-3 underestimate the maximum value, while the TALYS calculation give much better estimation for the high energy range above 40 MeV. Both codes give the maximum energy with acceptable accuracy.

\begin{figure}[h]
\includegraphics[scale=0.3]{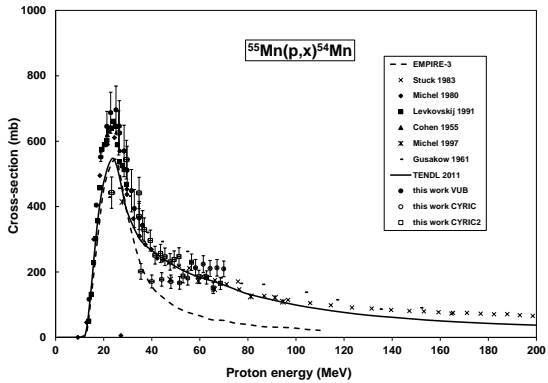}
\caption{Excitation function of the $^{55}$Mn(p,x)$^{54}$Mn  nuclear reactions compared with the literature data and the results of the theoretical model calculations}
\end{figure}

\subsection{The $^{55}$Mn(p,x)$^{52g}$Mn reaction}
\label{4.2}
The produced radioisotope $^{52g}$Mn has acceptably long half-life (5.6 d) to measure. It can be produced directly by the (p,3np) reaction, including the complex particle emission, but it can also be fed from the $^{52}$Fe mother, as well as in a small percentage from its own short-lived isomeric state 52mMn (T$_{1/2}$ = 21 min), so the given excitation function is cumulative. The results are presented in Fig. 4. Our new data support those of Stueck \cite{18} and Michel-1997 \cite{20}, the data of Michel-1980 \cite{19} are a bit larger around 40 MeV. Now the TENDL-2012 gives the better estimation, while EMPIRE-3 underestimates the experimental values above 60 MeV. The given maximum positions are correct by both codes, the TENDL-2012 shape is a better approximation of the experimental function.

\begin{figure}[h]
\includegraphics[scale=0.3]{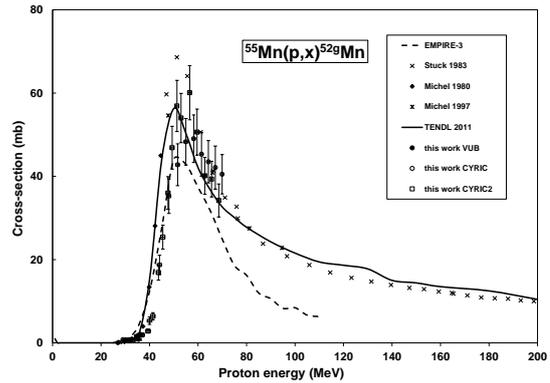}
\caption{Excitation function of the $^{55}$Mn(p,x)$^{52g}$Mn  nuclear reactions compared with the literature data and the results of the theoretical model calculations}
\end{figure}

\subsection{The $^{55}$Mn(p,x)$^{51}$Cr reaction}
\label{4.3}
The radioisotope $^{51}$Cr (T$_{1/2}$ = 27.7 d) is produced from manganese through the (p,3n2p) reactions most probably involving emission of complex particles, such as $\alpha$, d and $^3$He. It can also be fed from the parallel produced $^{51}$Mn.  According to Fig. 5 our new data show acceptable agreement with the earlier results of Stueck, Michel and Levkovskij \cite{18,19,20,21} in the covered energy range. Up to 50 MeV bombarding energy the TENDL-2012 and EMPIRE-3 results run approximately together, while over 50 MeV EMPIRE-3 gives better approximation and the TENDL-2012 strongly overestimates. 

\begin{figure}[h]
\includegraphics[scale=0.3]{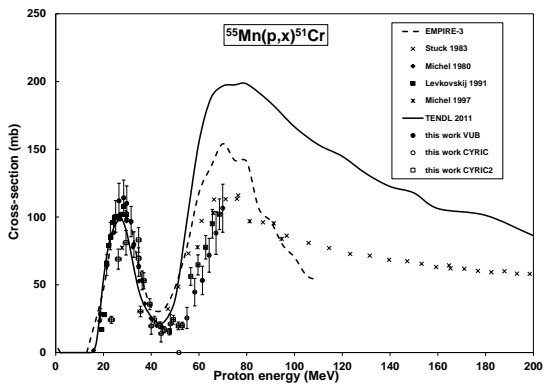}
\caption{Excitation function of the $^{55}$Mn(p,x)$^{51}$Cr  nuclear reactions compared with the literature data and the results of the theoretical model calculations}
\end{figure}

\subsection{The $^{55}$Mn(p,x)$^{48}$V reaction}
\label{4.4}
The radioisotope $^{48}$V (T$_{1/2}$ = 15.97 d) is produced from manganese through the emission of complex particles, as well as it could come from the decay of the mother $^{48}$Cr. Our new results are in acceptable agreement with the earlier result of Stueck and Michel \cite{18,20}(see Fig 6). The TENDL-2012 and EMPIRE-3 model calculations run parallel up to 60 MeV with a better approach of EMPIRE-3, but above this energy the estimation of EMPIRE-3 is better, although both models go far from the experimental values.

\begin{figure}[h]
\includegraphics[scale=0.3]{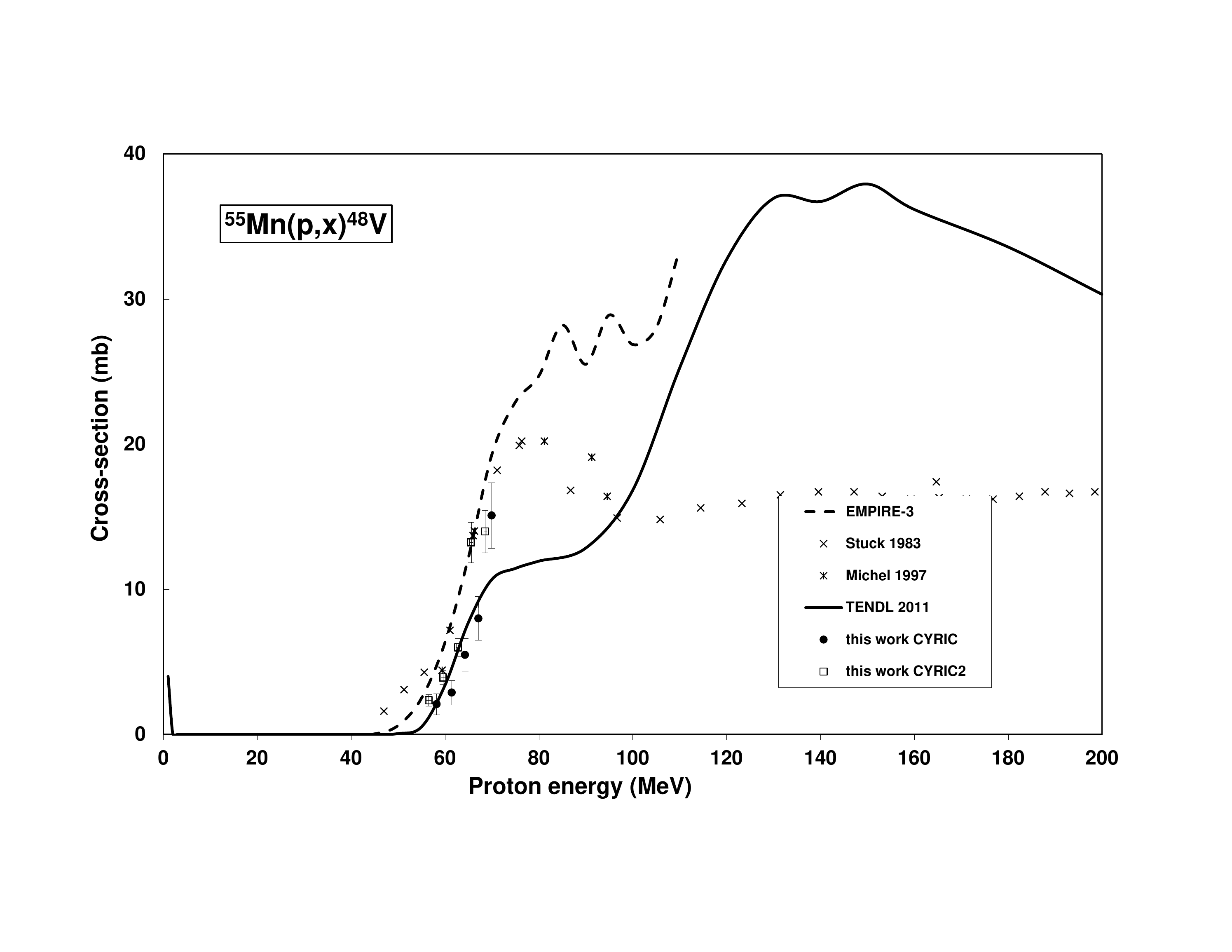}
\caption{Excitation function of the $^{55}$Mn(p,x)$^{48}$V  nuclear reactions compared with the literature data and the results of the theoretical model calculations}
\end{figure}

\begin{table*}[t]
\small
\caption{Measured cross-sections for ${}^{54}$Mn, ${}^{52g}$Mn, ${}^{51}$Cr and ${}^{48}$V}
\centering
\begin{center}
\begin{tabular}{|p{0.4in}|p{0.5in}|p{0.5in}|p{0.3in}|p{0.5in}|p{0.4in}|p{0.5in}|p{0.4in}|p{0.6in}|p{0.4in}|} \hline 
\multicolumn{2}{|c|}{\textbf{Bombarding}} & \multicolumn{2}{|c|}{\textbf{${}^{54}$Mn}} & \multicolumn{2}{|c|}{\textbf{${}^{52g}$Mn}} & \multicolumn{2}{|c|}{\textbf{${}^{51}$Cr}} & \multicolumn{2}{|c|}{\textbf{${}^{48}$V}} \\ \hline 
E & $\pm\delta$E & $\sigma$ & $\pm\delta\sigma$ &  $\sigma$ & $\pm\delta\sigma$ &  $\sigma$ & $\pm\delta\sigma$ & $\sigma$ & $\pm\delta\sigma$ \\ \hline 
\multicolumn{2}{|c|}{MeV} & \multicolumn{2}{|c|}{mb} & \multicolumn{2}{|c|}{mb} & \multicolumn{2}{|c|}{mb} & \multicolumn{2}{|c|}{mb} \\ \hline 
0.6 & 1.1 & ~ &  & ~ &  & ~ &  & ~ &  \\ \hline 
6.9 & 0.9 & ~ &  & ~ &  & ~ &  & ~ &  \\ \hline 
9.2 & 0.9 & ~ &  & ~ &  & ~ &  & ~ &  \\ \hline 
12.3 & 0.8 & ~ &  & ~ &  & ~ &  & ~ &  \\ \hline 
13.9 & 0.8 & 116.6 & 13.1 & ~ &  & 28.5 & 3.8 & ~ &  \\ \hline 
17.0 & 0.7 & 404.3 & 46.2 & ~ &  & 64.5 & 7.7 & ~ &  \\ \hline 
18.8 & 0.7 & 551.6 & 62.7 & ~ &  & 86.0 & 10.1 & ~ &  \\ \hline 
21.3 & 0.6 & 645.2 & 73.1 & ~ &  & 100.4 & 11.7 & ~ &  \\ \hline 
22.9 & 0.6 & 687.1 & 77.8 & ~ &  & 112.0 & 13.0 & ~ &  \\ \hline 
23.3 & 1.3 & 442.6 & 47.9 & ~ &  & 24.2 & 2.6 & ~ &  \\ \hline 
25.1 & 0.5 & 695.8 & 78.8 & ~ &  & 114.2 & 13.1 & ~ &  \\ \hline 
26.2 & 1.2 & 625.2 & 67.7 & ~ &  & 69.0 & 7.5 & ~ &  \\ \hline 
26.5 & 0.5 & 646.1 & 73.4 & 0.4 & 0.1 & 110.0 & 13.0 & ~ &  \\ \hline 
28.5 & 0.5 & 570.2 & 64.6 & 0.7 & 0.1 & 96.7 & 11.2 & ~ &  \\ \hline 
29.4 & 1.2 & 543.6 & 59.2 & 0.7 & 0.2 & 81.0 & 9.0 & ~ &  \\ \hline 
29.7 & 0.4 & 511.3 & 58.4 & 0.9 & 0.1 & 79.7 & 9.4 & ~ &  \\ \hline 
31.6 & 0.4 & 448.7 & 51.0 & 1.6 & 0.2 & 63.6 & 7.3 & ~ &  \\ \hline 
32.7 & 0.4 & 394.2 & 45.1 & 2.0 & 0.2 & 53.3 & 6.9 & ~ &  \\ \hline 
34.7 & 1.0 & 369.9 & 40.2 & 0.9 & 0.1 & 69.6 & 7.7 & ~ &  \\ \hline 
35.5 & 1.1 & 201.8 & 24.1 & 0.8 & 0.1 & 30.4 & 3.9 & ~ &  \\ \hline 
36.9 & 1.0 & 329.6 & 35.9 & 1.9 & 0.3 & 53.0 & 5.9 & ~ &  \\ \hline 
39.4 & 0.9 & 296.2 & 32.3 & 2.8 & 0.4 & 35.7 & 4.0 & ~ &  \\ \hline 
40.0 & 1.0 & 171.0 & 19.9 & 5.3 & 0.9 & 19.5 & 6.1 & ~ &  \\ \hline 
41.4 & 0.9 & 251.4 & 27.4 & 6.4 & 0.9 & 23.9 & 2.8 & ~ &  \\ \hline 
43.7 & 0.8 & 256.8 & 28.0 & 16.9 & 1.9 & 20.1 & 2.5 & ~ &  \\ \hline 
44.1 & 0.9 & 177.4 & 20.7 & 18.7 & 2.3 & 14.0 & 6.2 & ~ &  \\ \hline 
45.5 & 0.8 & 246.6 & 26.9 & 25.4 & 2.9 & 16.8 & 2.2 & ~ &  \\ \hline 
47.6 & 0.8 & 230.9 & 25.3 & 36.0 & 3.9 & 15.3 & 2.3 & ~ &  \\ \hline 
48.0 & 0.8 & 170.5 & 20.4 & 35.2 & 4.2 & 21.0 & 7.2 & ~ &  \\ \hline 
49.3 & 0.7 & 238.3 & 26.1 & 46.9 & 5.1 & 24.3 & 3.2 & ~ &  \\ \hline 
51.3 & 0.7 & 247.1 & 27.1 & 56.9 & 6.2 & 19.7 & 2.9 & ~ &  \\ \hline 
51.6 & 0.7 & 167.0 & 19.5 & 42.8 & 5.0 & ~ &  & ~ &  \\ \hline 
53.0 & 0.6 & 187.2 & 20.6 & 54.0 & 5.9 & 19.8 & 2.7 & ~ &  \\ \hline 
55.0 & 0.6 & 181.5 & 21.7 & 48.3 & 5.6 & 25.5 & 7.8 & ~ &  \\ \hline 
56.6 & 0.6 & 229.9 & 25.3 & 60.1 & 6.5 & 56.2 & 6.5 & 2.3 & 0.4 \\ \hline 
58.2 & 0.6 & 213.1 & 25.4 & 49.0 & 5.7 & 44.6 & 10.1 & 2.1 & 0.7 \\ \hline 
59.6 & 0.5 & 185.2 & 20.5 & 50.6 & 5.5 & 64.7 & 7.4 & 3.9 & 0.5 \\ \hline 
61.4 & 0.5 & 223.9 & 25.9 & 45.3 & 5.3 & 53.2 & 10.5 & 2.9 & 0.8 \\ \hline 
62.8 & 0.4 & 183.5 & 20.2 & 40.1 & 4.4 & 77.7 & 8.7 & 6.0 & 0.6 \\ \hline 
64.3 & 0.4 & 210.8 & 24.4 & 43.5 & 5.1 & 71.8 & 12.5 & 5.5 & 1.1 \\ \hline 
65.6 & 0.4 & 150.9 & 16.8 & 39.3 & 4.3 & 95.1 & 10.6 & 13.2 & 1.4 \\ \hline 
67.1 & 0.4 & 212.9 & 25.3 & 42.1 & 5.2 & 88.2 & 15.7 & 8.0 & 1.5 \\ \hline 
68.6 & 0.3 & 165.2 & 18.4 & 34.2 & 4.0 & 102.0 & 11.3 & 14.0 & 1.5 \\ \hline 
69.9 & 0.3 & 209.6 & 24.3 & 40.5 & 4.7 & 106.5 & 17.7 & 15.1 & 2.3 \\ \hline 
\end{tabular}

\end{center}
\end{table*}

\section{Physical yield}
\label{5}
By using our experimental results integral yield \cite{24} curves were calculated for the production of the observed radioisotopes. The results are presented in Fig. 7 in comparison with the few experimental points from the literature for some of the investigated isotopes \cite{25,26}. From Fig. 7 it is seen that the leading process up to 50 MeV is the $^{51}$Cr production as far as the activity regarded. Above this energy the produced activity of $^{52g}$Mn is strongly increasing. Experimental literature values exist only for 22 and 30 MeV for $^{51}$Cr and $^{54}$Mn respectively. The $^{51}$Cr values are strongly, while the $^{54}$Mn values are slightly under our calculated results.

\begin{figure}[h]
\includegraphics[scale=0.3]{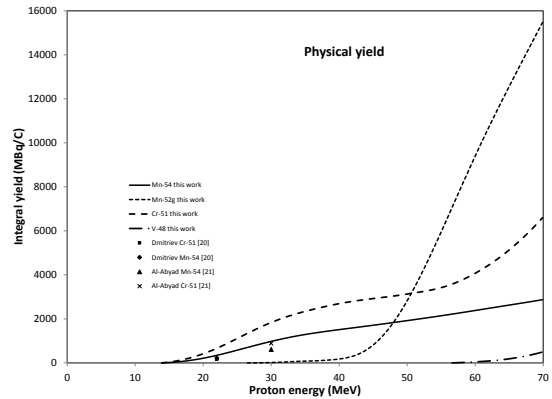}
\caption{Physical yield curves deduced from the experimental excitation functions for $^{54}$Mn, $^{52g}$Mn, $^{51}$Cr and $^{48}$V}
\end{figure}

\section{Activity distribution for thin layer activation (TLA)}
\label{6}
By using radioisotope tracing one can successfully monitor wear, corrosion and erosion processes in the swallow surface of materials \cite{27,28,29}.  For this purpose the $^{51}$Cr and $^{54}$Mn are the most proper choice from the presently investigated isotopes for manganese containing alloys. Both of the above radioisotopes have reasonable yield at lower energies and both have at least acceptable half-life. $^{51}$Cr can be used for quick processes, while $^{54}$Mn is more suitable for long lasting processes according to their half-life. All calculations are made with typical irradiation parameters of 1 hour irradiation time and 2 $\mu$A beam current. By $^{54}$Mn and perpendicular irradiation with 25.1 MeV the activity distribution was homogeneous within 198 $\mu$m within 1\% tolerance. The plotted results are calculated for 10 day cooling time (see Fig. 8). The 25.1 MeV bombarding energy was chosen as optimum energy for homogeneous activity distribution. If the surface layer to be investigated must be much swallower, very small irradiation angle should be chosen. The irradiation performed under 15$^o$ results in a more dense activity distribution in the surface up to 51.4 $\mu$m homogeneity. The advantage of the homogeneous activity distribution is that we need not know the distribution function, because the activity/unit depth is the same for the above referred depths (6 and 23 kBq/$\mu$m respectively). In the case of $^{51}$Cr the optimum bombarding energy for homogeneous distribution was 25 MeV and by perpendicular irradiation it resulted in 1\% homogeneity up to 170 $\mu$m. By 15$^o$ incidence irradiation the depth of homogeneity is only 44 $\mu$m (see Fig. 8). From Fig. 8 it is also seen that after only 10 day cooling time the specific activity of $^{51}$Cr is still higher than that of $^{54}$Mn, but because it has much sorter half-life as the $^{54}$Mn (see Table 2), it will quickly decay under that of $^{54}$Mn. While both of the above radioisotopes have local cross-section maxima in the investigated energy range, achieving homogeneous activity distribution is the most proper tool for wear measurements. In special cases, where the excitation function has no local maximum or even swallower investigation depth is required \cite{30}, one can produce linear activity distribution, by using lower bombarding energies. In this case the user performing the wear test must be provided with the exact activity distribution function \cite{31}.

\begin{figure}[h]
\includegraphics[scale=0.3]{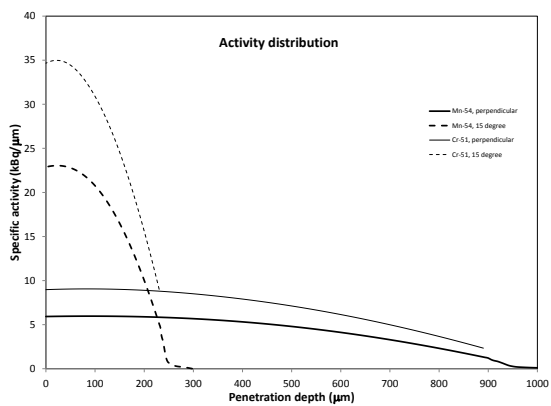}
\caption{Specific activity distribution (wear curve) for $^{54}$Mn and $^{51}$Cr performed under the following conditions: irradiation time 1 h; cooling time: 10 day; beam current: 2 $\mu$A; irradiation angles: 90$^o$ and 15$^o$; bombarding energies: 25.1 and 25 MeV respectively.}
\end{figure}

\section{7.	Conclusions}
\label{7}
The goal of the present work was: to re-measure and validate the earlier data by new measurements on the excitation functions on manganese; produce yield curves for potential users; provide data for improvement of theoretical model codes; as well as calculate activity distribution curves for thin layer activation. Our new measurements in most of the investigated cases show good agreement with the earlier literature data. The results of the two theoretical model codes TALYS (TENDL-2012) and EMPIRE-3 could reproduce the shape and the maximum position of the excitation functions quite well, but by quantitative estimation the EMPIRE-3 code performed better. The calculated yield curves make it possible to compare the performance the radioisotope production possibilities on manganese, as well as could help in designing experiments and/or industrial processes. 
As a main goal of this work specific activity versus penetration depth (so called wear curves) have also be constructed from our new experimental results for the most frequently used configurations in nuclear wear measurements (Fig. 8). It has been proven that both chosen radioisotopes are suitable for wear measurement by radioisotope tracing but for slightly different tasks, taking into account the time scale of the processes to be investigated.

\section{Acknowledgements}
\label{8}

This study was performed in the frame of the MTA-JSPS and MTA-FWO (Vlaanderen) collaboration programs. The authors thank the different research projects and their respective institutions for the practical help and providing the use of the facilities for this study.
 



\clearpage
\bibliographystyle{elsarticle-num}
\bibliography{Mnp}







\end{document}